# SketchPad$^{N-D}$: WYDIWYG Sculpting and Editing in High-Dimensional Space

Bing Wang, Puripant Ruchikachorn, and Klaus Mueller, *Senior Member, IEEE*

**Abstract**— High-dimensional data visualization has been attracting much attention. To fully test related software and algorithms, researchers require a diverse pool of data with known and desired features. Test data do not always provide this, or only partially. Here we propose the paradigm WYDIWYGS (What You Draw Is What You Get). Its embodiment, SketchPad$^{ND}$, is a tool that allows users to generate high-dimensional data in the same interface they also use for visualization. This provides for an immersive and direct data generation activity, and furthermore it also enables users to interactively edit and clean existing high-dimensional data from possible artifacts. SketchPad$^{ND}$ offers two visualization paradigms, one based on parallel coordinates and the other based on a relatively new framework using an N-D polygon to navigate in high-dimensional space. The first interface allows users to draw arbitrary profiles of probability density functions along each dimension axis and sketch shapes for data density and connections between adjacent dimensions. The second interface embraces the idea of sculpting. Users can carve data at arbitrary orientations and refine them wherever necessary. This guarantees that the data generated is truly high-dimensional. We demonstrate our tool's usefulness in real data visualization scenarios.

——————————— ◆ ———————————

## 1 INTRODUCTION

High-dimensional data analysis and visualization is useful in many applications and domains, and research on new techniques for this purpose has been progressing steadily. Designing new algorithm and software requires datasets with specific features for testing. However, real datasets are in limited supply and those that are available often – at least partially – lack the features needed for targeted evaluations. While synthetic datasets can be generated, this can be tedious and it often requires high programming skills to translate certain visual properties into statistical properties and vice versa. At the same time, data acquisition processes are never perfect and artifacts often arise that hamper data analysis routines such as clustering. Scientists require tools that allow them to edit their datasets but without disturbing true and important structures. Ideally such tools would operate in the same visual interface they already use to explore, analyze and reason with their data.

Here we present SketchPad$^{N-D}$, an interface for high-dimensional dataset generation and editing that is tightly integrated with high-dimensional data visualization. Users need not switch back and forth between data manipulation and visualization tools as they are combined into one interface. This provides better context for later iterations of the data generation process and facilitates a more streamlined workflow. As users are able to create datasets more quickly, they can explore and generate a larger number of these and possibly more complex ones. This in turn will favor the development of more robust algorithms and software for high-dimensional data analysis and visualization. Similarly, as users are able to edit data and artifacts more informed and thoroughly they will be able to make faster progress in their data analysis efforts.

Because most visualizations and common input peripherals are 2D, it is desirable to directly draw napkin sketches of a dataset into a selected visualization and so generate the specified data. Sketch-based interfaces are often used by novice users to create and manipulate data in many applications, but have never been explored

———————————————

- *Bing Wang, Puripant Ruchikachorn (PR), and Klaus Mueller (KM) are with the Visual Analytics and Imaging Laboratory, Computer Science Department, Stony Brook University, NY. Email: {wang12, pruchikachor, mueller}@cs.sunysb.edu. KM is also with SUNY Korea, and PR is also with Chulalongkorn Business School, Chulalongkorn University. The two first authors contributed equally to this work.*

in the domain of high-dimensional data. A recent system by Albuquerque et al. [1] allows users to draw 1D and 2D probability density functions (PDFs) or 2.5D probability distribution planes (PDPs) to define a dataset. But it is beyond argument that in the era of 'Big Data', we are encountering data with multivariate (ND) relationships that extend much beyond three intrinsic dimensions. Thus, we require suitable data design tools that can match this scope.

The interface we describe is based on two visualization techniques, namely parallel coordinates [11] and scatterplots – *dynamic* scatterplots to be precise [16]. Interactions in these visualizations update the data with immediate visual feedback. This gives the impression of WYSIWYG (What You See Is What You Get) and direct manipulation in popular WIMP (Windows, Icons, Menus, Pointer) user interfaces. Since our system supports ND generation and editing tasks – activities that are similar to *sculpting* in 3D but with a 2D drawing interface – we call our paradigm WY*D*ISWYG – What You *Draw* Is What You Get.

The parallel coordinates plot is a well-known visualization paradigm and flexible for arbitrary numbers of dimensions. In our SketchPad$^{N-D}$, while keeping the context of all dimensions in the same view at all times, users are able to draw a curve for the profile of a PDF in any dimension axis. To connect data between adjacent dimensions, users can sketch a bounding shape. Also, users can name each axis, set its minimum and maximum, and reorder it.

In addition to parallel coordinates, our system also features a data carving tool for generating and editing arbitrary multivariate data distributions via a scatterplot interface. This tool can operate either in a scatterplot matrix or with an interface for arbitrary projections [16]. As it is difficult for users to imagine what the effect of a specific carving operation done from one vantage point looks like from other vantage points, especially when the number of dimensions is larger than three, we provide simultaneous views from other vantage points of interest that reflect these shape changes.

Users can rely solely on one of the two tools, or alternatively start with one tool and use the other to further edit the current data. We demonstrate a number of examples that have followed this strategy.

Our paper is outlined as follows. Section 2 provides related work on both dataset generation and sketch-based interfaces. Section 3 presents an overview of our system, and Section 4 specifically describes our two proposed user interfaces. Section 5 presents some results generated with our system. The last section concludes this paper, lists some remaining challenges and points to future work.

## 2 RELATED WORK

One of our two user interfaces is based on parallel coordinates [11]. Parallel coordinates represents a point in Cartesian system coordinates as a polyline on parallel axes. The vertex position of a polyline on an axis denotes its value in the corresponding dimension. The other user interface is based on scatterplots and allows users to generate and carve data directly on the projection plane of the scatterplot. In addition to these two primary visualization techniques, other related works are grouped into two main subsections below.

### 2.1 High-dimensional Dataset Generation and Editing

Conventionally, a synthetic high-dimensional dataset is manually generated by running a small script specifying certain statistical features. Users who are not familiar with any general-purpose programming languages such as scientists in other fields may use a computing environment or language such as Matlab or R. To the best of our knowledge, there is no commonly used language specifically designed for complex high-dimensional dataset generation.

As synthetic data is useful in many fields, there are a number of automatic methods for specific applications. One of these is software testing based on constraints [7] using genetic algorithms [15][17]. Another is Google Refine [25] which is a tool to edit messy data, but it relies on data in text form and is not visual. The user also does not have a global view and the tool is not WYDIWYG. Baudel [23] provides a visual framework to edit large datasets. However, unlike our tool, theirs only allows users to edit existing data, and it is not possible to produce a new dataset from scratch. Albuquerque et al. [1] recently presented a system for high-dimensional dataset generation. While this tool is visual and interactive, it has two major limitations in the sense of high-dimensional space. Firstly, it allows users to define only 1D and 2D PDFs or at most PDPs which are on 2D sub-planes in 3D space. While this is certainly useful, it cannot capture the full gamut of realistic high-dimensional datasets which often have multivariate structures in subspaces of intrinsic dimensionality greater than three which cannot be reduced by affine transformations. These structures are also often not axis-aligned. We handle this problem by allowing users to navigate more freely in high-dimensional space and define distributions on any axis-aligned or non-axis aligned views of interest.

Secondly, the framework by Albuquerque et al. asks users to define all 1D PDFs and some 2D PDFs and PDPs, but it is often the case that one dimension is defined more than once. To avoid conflicts, they simply discard all distributions on a dimension if it has been previously defined. This, however, will inevitably result in the final generated dataset not satisfying all constraints. Our approach tackles this problem from another angle - we let users carve on existing data, not trying to generate it out of nothing. With this approach, defining one dimension multiple times is possible since users perform all manipulation directly in data space with the actual object in sight. As such, our system suggests a different workflow and tighter coupling between visualization and user interface.

Another recent work that could be useful for high-D data generation and editing is the iLAMP framework by Amorim et al. [2]. It allows users to insert new points into a 2D projection of the data which are then inversely mapped into high-D space. However, since the projections are obtained via a non-linear low-dimensional space embedding algorithm which only guarantees a locally linear mapping, the inserted points may not go exactly to their intended locations in high-D space. This is exemplified by a noticeable distortion of a drawn and inversely mapped shape when it is mapped back from high-D to 2D space. Therefore this interface is not WYDIWYG, but it was not conceived to be used for sketching anyhow – rather it is being used to insert seed points for optimization algorithms.

Finally, a framework [3] has been recently proposed that uses a parallel coordinate interface for data generation. Its publication coincided with that of our preliminary work [19] but unlike ours, their interface does not operate within a sketching paradigm.

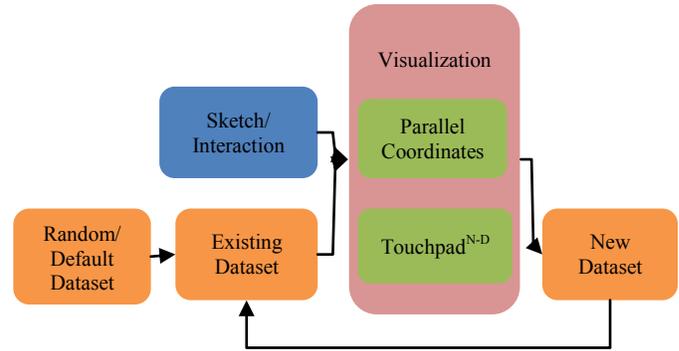

Fig. 1. Overview of our high-dimensional dataset generation workflow, along with the proposed user interfaces

### 2.2 Sketch-based Interfaces

Sketch-based interfaces have been explored in many applications, especially those with large numbers of novice users. Sketch-based interfaces interpret gestures from common 2D input devices such as mouse, touch screen, and stylus or pen as application-specific outputs. For example, there are gestures for entering music notations [8], composing music [12], solving mathematical problems [13][22], and generating 3D molecular structures [20]. Sketch-based modeling is a specific application in computer graphics that converts 2D shapes and gestures into 3D models [21][10].

There is some recent research on sketch-based visualization as a casual communication tool [4][5][6]. But the end product of these applications is a mock-up of a real visualization, not a dataset. For data manipulation, basic interactions are provided to select or brush data points but these are usually limited to trivial visualizations in a 2D canvas such as a scatterplot. To the best of our knowledge there is no previous work on a sketch-based interface for high-dimensional dataset generation.

## 3 OVERVIEW

Our proposed workflow for data generation is shown in Figure 1. It differs from a normal visualization in that interactions change the data, which in turn are updated and displayed for further editing. It also differs from a normal data generation procedure where all data manipulation occurs before visualization. Our workflow and its concept of tying visualization with a user interface for dataset generation and editing is collectively called SketchPad$^{N-D}$. In our system, users can start with an existing dataset, a random dataset or from scratch, and then modify this initial data object using our two tools. The resulting dataset can then be fed into the system again for further editing and processing.

Users are free to use either of our two editing tools in succession. The data format is shared among our editing tools and so each iteration in the data generation process does not have to be visualized and edited by the same technique. Even though our tools can be used independently, we chose two visualization paradigms that complement each other. Other paradigms with specifically designed interactions can be easily added into the workflow.

## 4 USER INTERFACES

Our proposed user interfaces are tightly coupled to visualizations in order to provide a better linkage between data visualization and data generation. This should give an impression of natural ease of use due to the psychological proximity of the generated dataset and its visual representation. Many research efforts suggest ways to visualize data but not how to create and edit them. This one-way workflow from data to visualization has been abundantly explored in the literature but not the other way around.

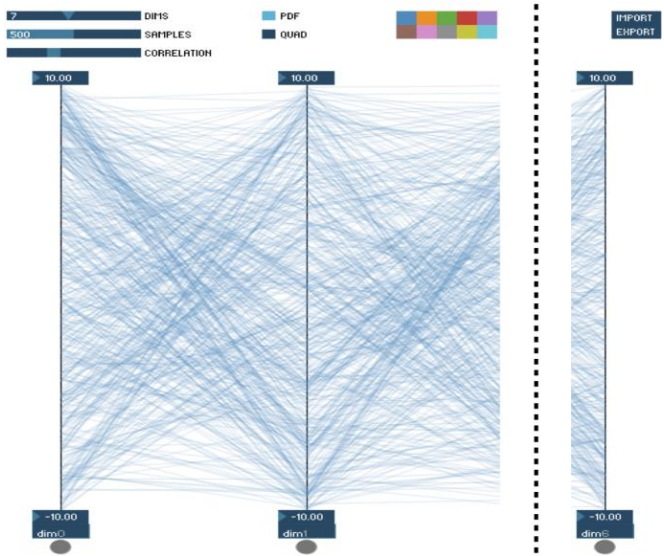

Fig. 2. The interface for high-dimensional dataset generation using the parallel coordinates interface.

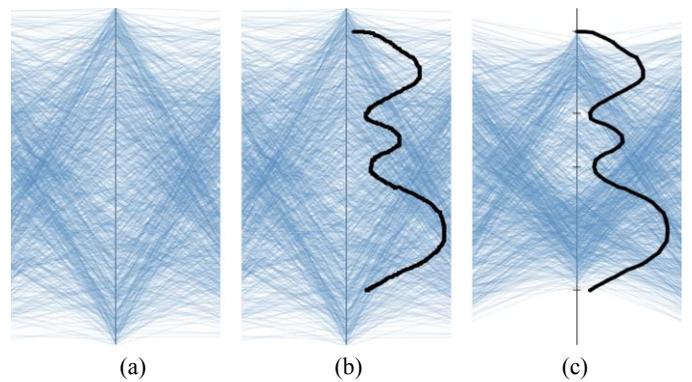

Fig. 3. (a) A uniformly distributed 1D dataset, (b) a PDF sketch curve on its axis, (c) generated data according to the resampled and snapped PDF while data in adjacent dimensions remain uniformly distributed.

### 4.1 Sketching on Parallel Coordinates

The main user interface for this part is an extension of standard parallel coordinates. As shown in Figure 2, on the top left, there are three sliders to adjust the number of dimensions, the number of generated samples per cluster, and the correlation per cluster. Next to them, there is a group of radio buttons for two modes for sketching.

There is also an array of color patches where users can select the pen color that represents each cluster. In this prototype, there is a limit to at most 10 clusters per dataset but this constraint is easily expandable to an arbitrary number of clusters as the screen real estate allows. When a color is selected, its corresponding number of samples per cluster is reflected in the slider.

In the top right corner, there are two buttons for importing and exporting data in space-delimited format. A random dataset, which is uniformly distributed in each dimension, is loaded by default. Users can import any existing dataset in this format. Also, users can export the dataset to visualize or edit it further in any other application. One possible workflow is to export the dataset from a quick sketch and edit it in the scatterplot interface (see Section 4.2).

From top to bottom, each axis of the parallel coordinates has two number boxes and one textbox for displaying and setting the maximum, minimum, and name of the dimension, respectively. To reverse the range of a dimension, users can switch its maximum and minimum values. Also, users can reorder dimensions by dragging a handle below each axis and dropping it between any pair of adjacent axes to move the selected dimension.

The parallel coordinates are displayed prominently in the middle of the screen where users primarily interact. A uniformly distributed cluster of 500 samples and 7 dimensions in the range of -10.0 to 10.0 is prepopulated. When users draw via some predefined gestures, the data is updated interactively in direct manipulation. The two modes of sketches selected by the radio buttons mentioned above are explained in detail in the following subsections.

#### 4.1.1 Probability Density Function (PDF) Sketch

In this mode, users can freely draw an arbitrary curve along any axis to specify how data should distribute in the corresponding dimension. This curve is then interpreted as a PDF of the data in that dimension, with associated skewness, kurtosis, and other properties.

Figure 3 illustrates this process. Figure 3a shows the initial random point distribution ready to be shaped. Next, in Figure 3b, the user sketches a rough curve which is resampled along its length to simplify later computation and smooth appearance. This curve is matched to the dimension whose parallel coordinate axis is the closest to the leftmost point. Because each dimension can have at most one distribution and hence one PDF, the new sketch overwrites, if any, the existing one. Now, to interpret the resampled sketch as a PDF, the line is shifted to set its leftmost point as zero and its end points are extended to the baseline. A function of the normalized distance between the leftmost baseline and each point in the extended curve satisfies the definitions of a continuous PDF i.e. it is equal or greater than zero, its limits at positive and negative infinity are zero (due to baseline-extended endpoints), and its overall integral is one (due to normalization).

Then the continuous PDF is sampled to create a discrete cumulative distribution function (CDF) for data generation purposes. Following inverse transform sampling, a data value in a PDF-specified dimension is generated by finding the index of a uniformly distributed random variable in [0, 1) from the discrete CPF. The tent filter is applied later to linearly interpolate discrete indices [1]. As these PDF sketches are independent to each other, only the data at their own dimensions are updated (see Figure 3c). Clusters are indiscriminately affected by this distribution.

#### 4.1.2 Data Connection Quadrilateral

Drawing two univariate PDF curves on adjacent parallel coordinate axes is insufficient to concretely define the bivariate distribution of these two variables. Given $N$ points there are in fact $N!$ such distributions. It is effectively only the user who is to decide which of these many distribution shapes to choose and we provide a set of easy-to-use design tools for this purpose. The design elements we provide adhere to the common patterns often observed in parallel coordinates: a bowtie for negative correlations and a trapezoid for positive correlations.

In our interface users can draw a quadrilateral between two adjacent axes to specify a data connection between the two corresponding dimensions at particular ranges. Each click adds a vertex to the quadrilateral and four clicks form a shape that can be simple or complex (self-intersecting).

As shown in Figure 4, the vertices of a quadrilateral are extended along their most horizontal sides to snap to the closest axes. A simple quadrilateral becomes a trapezoid and a complex quadrilateral becomes a bowtie with one pair of vertically parallel sides. The quadrilateral is rendered in translucent cluster color so the generated data are shown through.

The vertical sides on both axes specify affected ranges of the two corresponding dimensions. A trapezoid and a bowtie represent a direct and an inverse relationship, respectively. Values in the left range of a trapezoid tend to match directly with the ones in the right range from top to bottom and values in the left range of a bowtie do so inversely.

A multivariate dataset is typically developed from the leftmost to the rightmost dimension. Derived from the generated value for each data sample at a dimension and the input quadrilateral, a temporary distribution for the next (neighboring) dimension is created and used to generate values for that dimension in this sample. Based on rejection sampling, accepted are only values within the range of their temporary distributions. This step repeats through all dimensions for one sample.

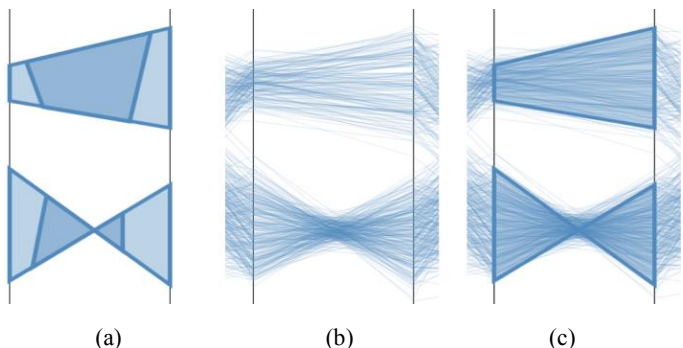

(a) (b) (c)

Fig. 4. (a) Sketched input quadrilaterals (solid blue) between a pair of adjacent dimensions and after snapping their vertices to the axes (faint blue), (b) the generated data, and (c) their superimposition.

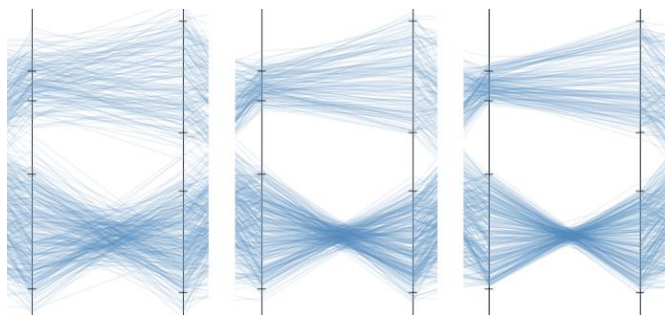

Fig. 5. From left to right: generated data of the same trapezoid and bowtie with different correlations (greater towards the right).

As discussed earlier, in parallel coordinates, negative correlations are signified by bowtie shapes while positive correlations give rise to trapezoids. Apart from sign, another important parameter is correlation strength. In parallel coordinates, the strength of a correlation is visualized as the range in which line crossings occur within a quadrilateral or bowtie. A perfectly positive correlation would have the same ordering of lines on both adjacent axes while a lesser amount of correlation will have lines that cross within a small neighborhood. We allow users to modify the correlation strength within a cluster by adjusting a slider. Setting this slider changes the size of a small sliding window within the shape that determines the range of possible values in the next dimension -- the lower the correlation, the larger this window. The window size is configured as a percentage of the range of the quadrilateral on the corresponding axis. An example is shown in Figure 5, where the greater the correlation, the narrower the sliding window and the less the number of line crossings is for the generated data of a trapezoid. The same slider is also used for negative correlation signified by the bowtie.

A quadrilateral affects only samples of the same color so all independent clusters can be generated despite their overlapped domains. For convenience, drawing a quadrilateral in a color that has zero number of samples sets it to a default number, 500 samples per cluster.

### 4.1.3 PDF Sketch and Data Connection Quadrilateral

Sketching PDFs and drawing quadrilaterals independently can lead to data inconsistencies that have no meaningful clusters as is shown in Figure 6a. Our interface prevents this by adding positional constraints to locations where neighboring dimensions already have existing sketches or shapes. More specifically, a PDF sketch on one dimension axis restricts any new quadrilaterals next to it and a quadrilateral on one dimension pair restricts PDF sketches on the two dimensions and any quadrilaterals next to both dimensions of the pair.

In Figure 6, the local leftmost points in a PDF sketch create ticks onto the axis for a quadrilateral to snap to when its extended vertices are in proximity. Likewise, a quadrilateral creates four ticks from its extended vertices for other quadrilaterals to snap to. Finally, a PDF sketch drawn on an axis that has quadrilaterals will be clipped to the ranges of those quadrilaterals.

### 4.2 Sketching on Scatterplots

A second paradigm in high-dimensional data visualization is the scatterplot – a projection of all data points onto a two-dimensional plane that is either aligned with two of the data axes or in general position. Our interface is shown in Figure 7. It consists of the following components:

**Scatterplot display:** This window shows the projected N-D data and serves as the main editing canvas. The data are projected as points into the 2D basis formed by the *projection plane axis (PPA) vectors*. These two orthogonal N-D vectors define the *x*- and *y*-axes of the resulting scatterplot display. The projected data axes can be optionally visualized directly in the scatterplot display or in isolation in the **data axes vector display**. The **vector component bar chart display** conveys more information about the dataset. The top two charts show the *x* and *y* components of the current PPA vectors while the bottom chart shows the PCA spectrum of the data.

**N-D touchpad polygon:** This is the interface by which the orientation of the PPA vector basis can be interactively configured, which in turn determines the projective view onto the data. Each polygon vertex represents a data dimension vector. The PPA vectors are determined by the positions of the red and blue points in the polygon's interior using generalized barycentric interpolation [14]. The **touchpad polygon control panel** lets the user switch between moving the red point (PPA x-axis) or the blue point (PPA y-axis).

**Scatterplot display controls:** This panel allows users to change the scatterplot's point size, zoom in or out, view the points in a 2D or 3D display, and see the moving trajectory as motion blur for better shape perception in motion parallax.

**Sketching controls** enable users to input the number of dimensions, the number of points, choose to start from scratch or with an existing dataset, and modify the points in axis-aligned or non-axis aligned fashion. **Brushing controls** let the user switch between brushing and erasing mode, choose brush size, brush density and brush color.

**Scatterplot matrix (SPLOM) window:** This is a separate window that is used when editing the distributions in axis-aligned mode (see Section 4.2.1).

**Scatterplot views window:** This is another separate window that is used when editing the distributions in non-axis aligned mode (see Section 4.2.2)

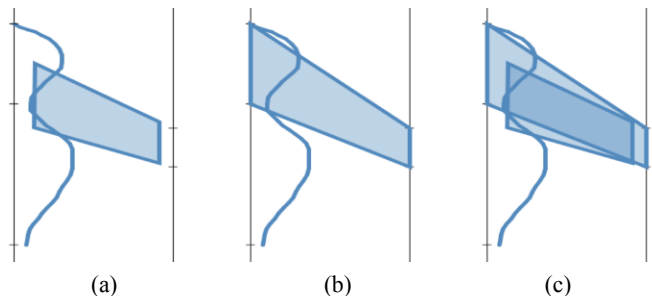

(a) (b) (c)

Fig. 6. (a) An example of an inconsistent PDF sketch and data connection quadrilateral, (b) the meaningful snapped version, and (c) their superimposition.

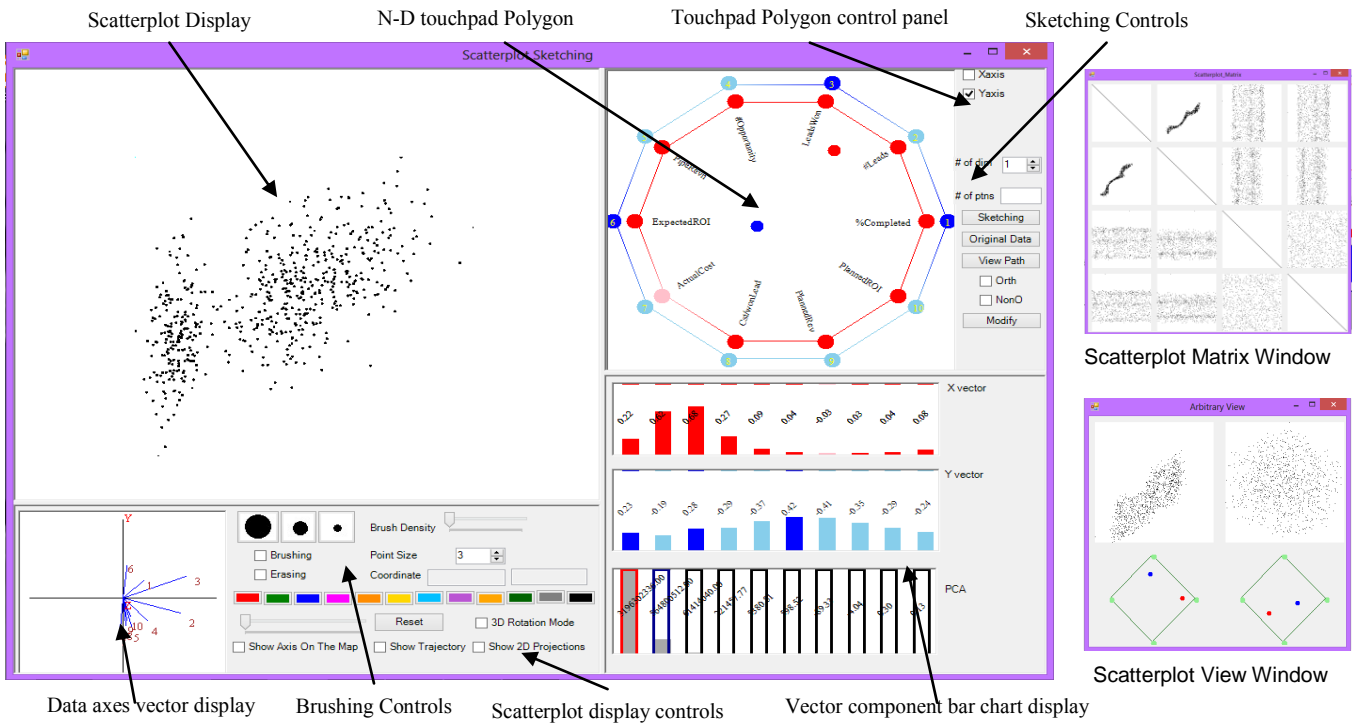

Fig.7. Scatterplot sketching interface

**Distribution designer:** Users design an initial distribution using this window (see Figure 9 and explanatory text in Section 4.2.1, Step 1).

Readers interested in the scatterplot display and how it is interactively controlled using the N-D touchpad polygon are referred to [16].

The design process begins with the user drawing an initial 2D distribution shape using the distribution designer. This 2D distribution is conceptually due to a collection of N-D points projecting into this shape. Initially these points would be randomly distributed in the other $N$-2 dimensions. Now we can pick another scatterplot projection and carve the 2D projected point cloud into one that we like to see from this orientation, under the constraint of the first drawing. We can repeat this for other projections and so on. This carving is essentially a subtraction of points from the current N-D distribution. However, a fundamental problem related to the nature of high-dimensional space is that the carving may subtract points in undesired locations in other projections. Therefore we need to replenish the N-D distribution every once in a while to satisfy all prior constraints, i.e. the shapes drawn and carved so far. Our proposed algorithm is presented in Figure 8. We will demonstrate this algorithm first for the axis-aligned case and then for the non-axis aligned case. When interacting, users can switch among these modes at will.

### 4.2.1 Interactions with Axis-aligned Scatterplots

Axis-aligned interactions make use of the SPLOM window for high-D space visualization. While this does allow for the generation (and editing) of distributions of any dimensionality, it requires them to be axis-aligned -- the non-axis aligned interface discussed in Section 4.2.2 removes this constraint. As outlined in Figure 8, the point generation stage consists of two operations: distribution painting and backprojection. The sculpting supports two operations as well: distribution carving and repair. Editing operations use the same framework – just now a distribution already exists and does not have to be initialized. Editing an existing dataset is a good idea even when data generation is the goal – one does not have to start from scratch.

We now describe each of these four operations in turn, using Figure 10 as a running example.

**Step 1: Distribution Painting.** This activity occurs in the distribution designer window (see Figure 9). The user first selects the desired axis-aligned view by placing the two PPA vector points in the N-D touchpad polygon onto the respective vertices. He then uses the sketching brush to draw the boundary of the distribution (Figure 9a) and then the centerline (Figure 9b). The center line is where the densest points should appear while the boundary constrains the range of the points. Next he uses the distribution profile designer to define the profile of the distribution (Figure 9c). This fills the drawn shape with a point distribution which looks fairly regular. The user now has the option to paint additional points to make this distribution look more natural, i.e., less regular using a paint brush. The density of these drawn points can be controlled using a slider (Figure 9d). Alternatively, points can also be removed using a fuzzy eraser, which strength is also configured by the density slider. The resulting 2D distribution is then converted into a probability map that governs how N-D points will project onto this projection plane. To ensure a

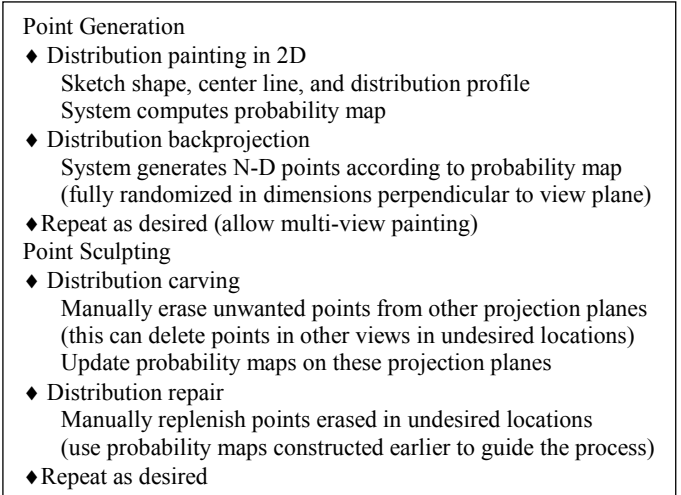

Point Generation
- ♦ Distribution painting in 2D
  Sketch shape, center line, and distribution profile
  System computes probability map
- ♦ Distribution backprojection
  System generates N-D points according to probability map
  (fully randomized in dimensions perpendicular to view plane)
- ♦Repeat as desired (allow multi-view painting)

Point Sculpting
- ♦ Distribution carving
  Manually erase unwanted points from other projection planes
  (this can delete points in other views in undesired locations)
  Update probability maps on these projection planes
- ♦ Distribution repair
  Manually replenish points erased in undesired locations
  (use probability maps constructed earlier to guide the process)
- ♦Repeat as desired

Fig. 8. Scatterplot based data generation algorithm

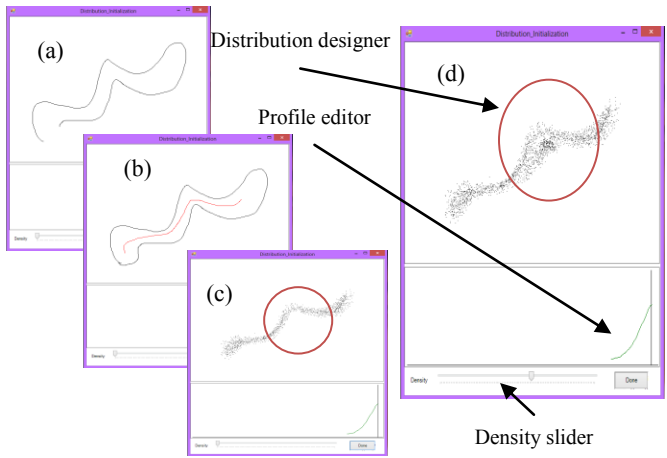

Fig. 9. Scatterplot sketching process. (a) Draw the boundary of the distribution. (b) Draw the center of the distribution. (c) Draw the profile of the distribution and the backprojected points. (d) Use the density bar to control the brush density and add more points.

sufficiently smooth probability map, we could optionally filter this density map with a 2D Gaussian function before backprojection. Figure 10a shows our running example in the SPLOM window – we sketched the shapes and centerlines of four clusters.

**Step 2: Distribution Backprojection.** Once the probability map has been constructed, the system uses it to generate $P$ N-D points where $P$ can be specified by the user. For the axis-aligned case, since all other projection planes at this initial configuration have a projection probability map with uniform distribution we can randomize the coordinates of all other $N$-2 dimensions. The result of this step is a set of $P$ N-D data points with all coordinates defined. Figure 10b shows the result of this step for a 4-D scatterplot matrix [9], composed of dimensions *x1, x2, x3,* and *x4*. Since the painting occurred in the *x1-x2* plane (which is now fully defined), the *x3-x4* projection has still a random point distribution. All other projections are partially defined since they either include dimension *x1* or *x2*.

Different distributions drawn in step 1 can be treated as different clusters and colored differently, as shown in the scatterplot matrix window of Figure 10b. The coloring clearly distinguishes between each drawn distribution and lets the user quickly see the points belonging to each (sub-) cluster. For distributions that may be too complicated to define at once or span more than one view, it helps to decompose them into simpler distributions and use the color coding as a guide to work on them one by one. For example, the final cluster shown in Figure 10i (using four views in the scatterplot display) spans the entire 4D space but is composed of four different-colored sub-clusters each spanning just one dimension.

To create even more complex clusters the user may repeat the first two steps as often as desired by placing the PPA vector points on other vertices in the N-D touchpad polygon to select another view and paint on it. In this operation, all previously generated points are shown as inactive background labeled in gray color (Figure 12b). This multi-view painting mechanism allows users to define highly multivariate clusters. We found that these two levels of decompositions can help greatly in comprehending the N-D sculpting task.

This point generation procedure just described provides us with a set of initial points to further work on – especially in those views that have at least one dimension not part of the painting process. Users can now sculpt on those dimensions to finish the data generation task.

**Step 3: Distribution Carving.** Figure 10c visualizes the carving effects on all projections simultaneously in the scatterplot matrix window. In this figure the carving occurred for cluster 1 (black) in the *x1-x3* plane – all other three clusters are inactive now and are shown in gray. The user carves the points in the desired locations directly in the scatterplot display. The carving tool can also work as a fuzzy eraser such that every swipe only removes a random subset of the covered N-D points. Similar to the brush used in the data generation step, users can choose from three different eraser sizes. A large eraser allows quick point removal while a small eraser refines details. At every eraser location the algorithm picks a random set of points that project into a small box around it – the size of this box can be user specified – and these points will be removed from the location's current list of N-D points. Following, the updated distribution is stored in this projection's probability map. However, note that removing points from this projection's list will also remove these points from any other projections in the scatterplot matrix and update (scale down) their respective probability maps. This can lead to undesired effects and so the final (repair) step is required.

**Step 4: Distribution Replenishing/Repair.** As just mentioned, erasing/carving points in one view may delete points in other views in undesired places, and so we will have to bring these points back. Further, the repeated erasing of points will deplete the N-D distribution in general. Figure 10c shows an example of this effect – the distribution of cluster 1 in the *x1-x2* plane is much weaker than in Figure 10b. We can repair/replenish the N-D distribution in two ways: (1) automatically by comparing the original projection probability maps with the current projections, and (2) user-driven by allowing the user to paint the missing points back on. The latter facility can also be used for general editing. In either mode, we need to make sure that we add points only in desired places. Given the current projection location subject to repair, we randomize the coordinates of the dimensions whose distributions have not been defined yet. For the other dimensions (excluding the two forming the current projection) – those for which the distributions have already been defined -- we compute their joint probability map and from it generate their coordinates. Figure 10d shows an example where we observe that the distribution of cluster 1 has gained back the strength it had in Figure 10b.

**Step 5 and on: Repeat.** The user can pick any projection and update its probability map by carving or replenishing. This will activate the procedures described in detail in the previous steps. Figure 10 e-h show a few more such operations for our demonstration dataset. The result is visualized in Figure 10i using four non-axis aligned scatterplot projections. The generated data resembles a 4D snake-like structure which could not be constructed with existing tools that operate in 2.5D space.

### 4.2.2 Interactions with Non-axis Aligned Scatterplots

Non-axis aligned scatterplots provide additional freedom in cluster design but they require a more sophisticated navigation interface. A non-axis aligned view is selected by placing the PPA vector points inside the N-D touchpad polygon (see Figure 11a). We now describe the five data operations in turn.

**Step 1: Distribution Painting:** This is similar to the axis-aligned case, just now the selected view is non-axis aligned.

**Step 2: Distribution Backprojection:** This process is significantly more complicated than in the axis-aligned case. We first build a new coordinate system spanning the same data space using the Gram-Schmidt orthonormalization process. It takes $N$ linearly independent vectors and produces $N$ orthonormal vectors spanning the same N-D space. Let $<x_1, x_2>$ denote the inner product of two vectors $x_1$ and $x_2$, $\|y\|$ denotes the length of vector y. Let the projection of vector $x_2$ onto vector $x_1$ be $proj_{x_1}(x_2) = x_1 \frac{<x_1,x_2>}{<x_1,x_1>}$. Gram-Schmidt starts with $N$ linearly independent N-D vectors $\{x_1, x_2, x_3, ..., x_N\}$ and performs the following calculations:

$$y_1 = x_1, \qquad\qquad\qquad\qquad\qquad e_1 = y_1/\|y_1\|$$
$$y_2 = x_2 - proj_{y_1}(x_2), \qquad\qquad e_2 = y_2/\|y_2\|$$
$$y_3 = x_3 - proj_{y_1}(x_3) - proj_{y_2}(x_3), \quad e_3 = y_3/\|y_3\|$$
$$\vdots$$

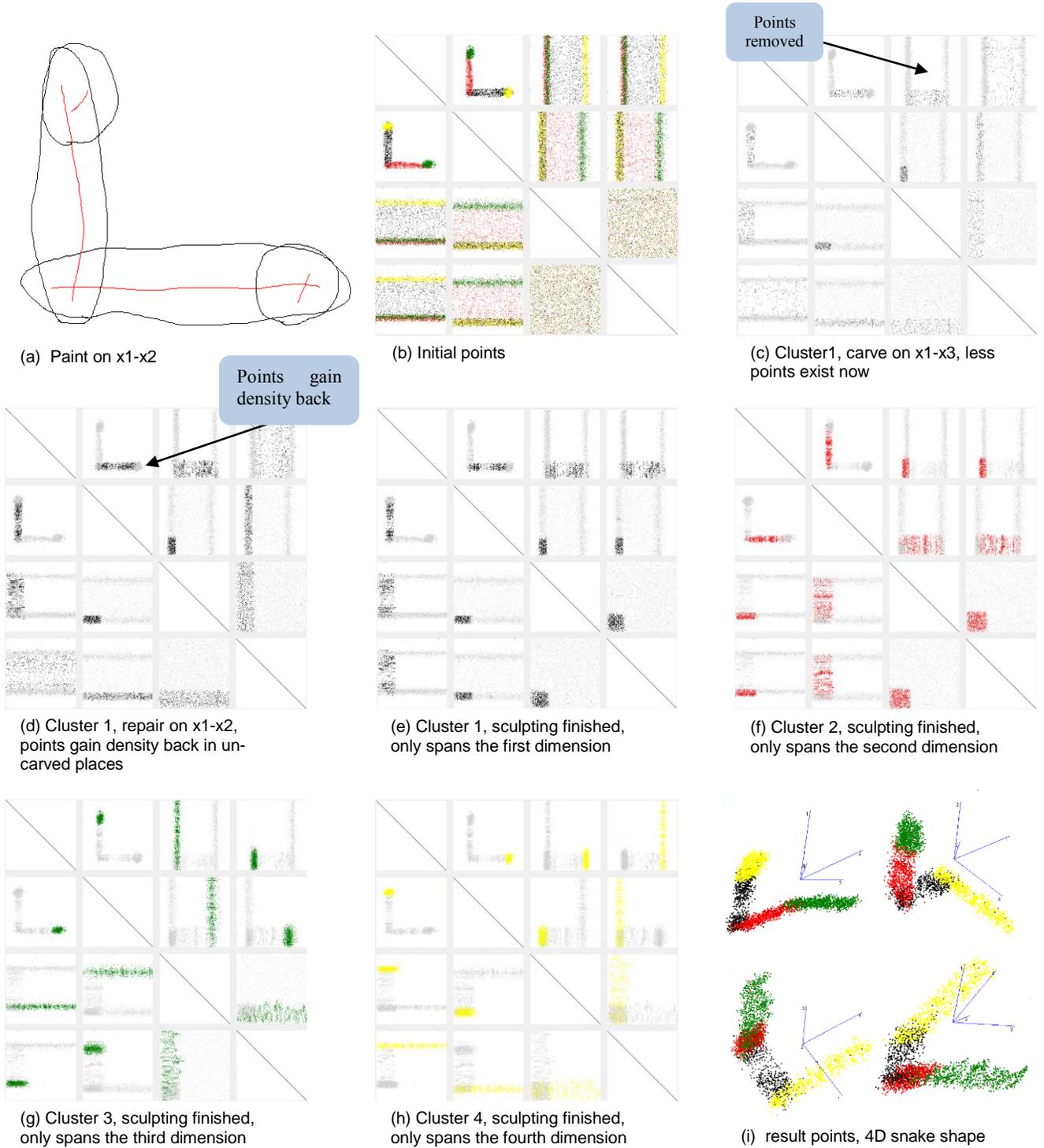

Fig. 10. An illustrative example of the scatterplot based data generation algorithm using axis-aligned views in the SPLOM window (panel (i) shows four views in the scatterplot window)

$$y_N = x_N - \sum_{j=1}^{N-1} proj_{y_j}(x_N), \qquad e_N = y_N/\|y_N\|$$

The resulting vector set $\{e_1, e_2, e_3 ..., e_N\}$ is the orthonormal set. In our case, we randomly generate ($N$-2) N-D vectors – we keep the two user selected vectors in order to preserve the shape the user painted. We also make sure that those vectors are linearly independent. Following we apply the Gram-Schmidt process on the $N$ vectors. We treat the new orthonormal vectors as general unit base vectors and generate data using the same procedure as used in the axis-aligned case. Since the projection (painting) plane is not axis aligned, the generated point coordinates are described in the rotated space. We can calculate their true data axis-aligned coordinates by multiplying them by the rotation matrix formed by the orthogonal basis vectors described above.

**Step 3: Distribution Carving:** As with the SPLOM in the axis-aligned case, we require a set of simultaneous views to enable users gain a more comprehensive understanding of the effect of the current carving. In the non-axis aligned case discussed here these views can be arbitrarily chosen by the user and typically hold views on certain structures the user would like to maintain. Then, any carving interaction in one view is immediately updated in all other views. Undesired effects can be undone by pressing an undo button and similar to an actual drawing process done on paper, we also provide a tool for erasing any unsatisfying mistakes.

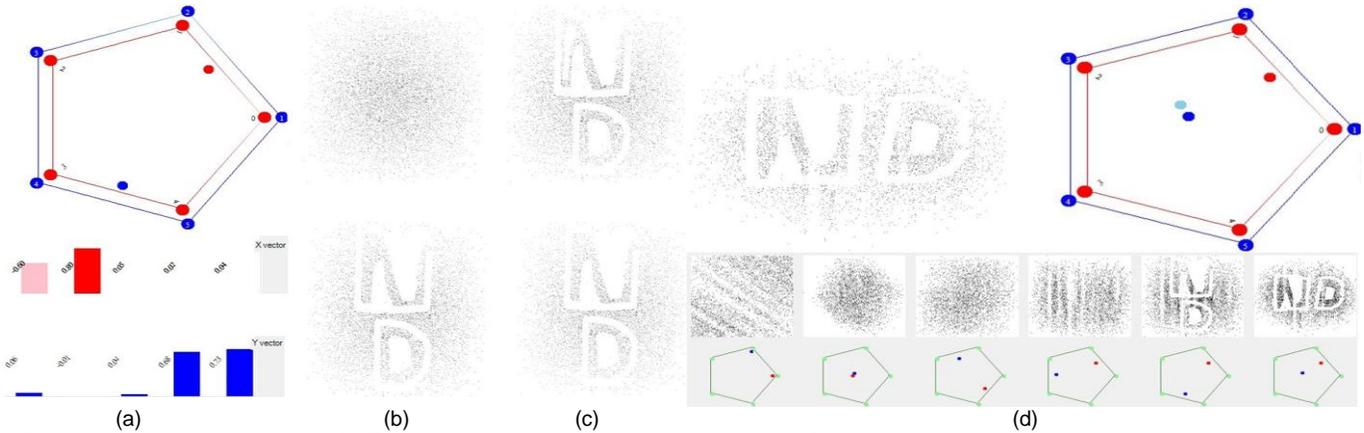

Fig.11. Non-axis aligned carving. (a) Top: the composition of the touchpad polygon. Bottom: the weighting of each dimension for the x and y PPAs. (b)(c) Carve and erase on one non-axis-aligned plane. (d) Select another plane to carve another shape. Note that the carved shapes can only be seen on those two selected views.

**Step 4: Distribution Replenishing/Repair:** This is more problematic than for the axis-aligned case in which projections were mutually orthogonal. In the non-axis aligned case, unless two projections are proven to be orthogonal to each other, any view selected will partially overlap with another. Hence, when carving out points in one view, this effect cannot be repaired in another view using the above method. Our current solution is to allow the user to repair by bringing all points back, even if they have been carved out in the first view.

**Step 5 and on: Repeat.** This is similar to the axis-aligned case.

### 4.2.3 Application Example for Non-axis Aligned Scatterplots

Figure 11 demonstrates an example using an initial 5D Gaussian distribution. Our result dataset displays two characters 'N' and 'D' on the projection plane of which the x-axis is close to the first and second dimensions while the y-axis is close to the fourth and fifth dimensions. Figure 11a shows the configuration of our polygonal touchpad. We move the red point close to the first and second dimensions and the blue point close to the fourth and fifth dimensions. The two bar charts below show the different dimension interpolation weights for the x and y-axis. We can clearly see that the first and second dimensions dominate the composition of the x-axis while the fourth and fifth dimensions have the largest value for y-axis. Figure 11b shows the two multivariate scatterplots of the data on the previously selected projection plane, before and after carving. The top scatterplot shows a point cloud while the bottom displays the two characters after carving. This pattern can be observed only on this plane and it may not show any meaningful patterns on any other plane. Figure 11c is a case in which the eraser comes in handy. The top scatterplot shows that a part in the 'N' is accidentally brushed away. Note the hollow portion on the left part of 'N'. Like in any painting tool, we can change to 'erasing' mode and use the eraser to bring those points back as shown in the bottom scatterplot. With the help of the brush and eraser users can create any pattern on any projection plane.

Next, the user may want to carve the existing set of characters such that they are also visible from another projection orientation. Carving on another projection plane may compromise the previous patterns so the user needs to be more careful. As is shown in Figure 11d, the user navigates to the second plane using the touchpad, adds this projection to the row of multiple views and starts carving. He can easily monitor the changes on the previous projection (the fifth) and make sure the ongoing carving does not adversely affect the overall shape of the data there.

We note that changing the carving plane can extend the intrinsic dimensionality of the resulting dataset to the number of dimension of the touchpad polygon. Our present example has a 5D polygon but there are no limits on how many vertices such a polygon can have. We note, however, that the ordering of the vertices is important with respect to the subspace of the data that can be reached [16]. We may initialize or reconfigure this vertex ordering using the parallel coordinate interface, but our navigation polygon also allows users to interchange, add, or remove vertices directly. Likewise, when a SPLOM is used and the dimensionality of the data grows high, the user may choose the dimensions composing the SPLOM using a selection interface.

### 4.2.4 Active and Inactive Points

The navigation polygon and the multiple views help users to easily carve data globally but not locally. Because of the high-dimensional space, a subset of data is cluttered with other data once the plane changes. Our interface also supports the concept of active and inactive points to provide the user complete control over selecting an active subset of data points for editing. Users can assign different point sets to clusters and control their status (active/non-active) via a check-box interface. In this case only the active points can be manipulated via the brushing/eraser tools. For example, all points marked in light gray in Figure 10 are currently inactive and will be insensitive to any brush manipulation.

### 4.2.5 Dataset Generation for Clustering Algorithm Testing

As mentioned, one of the most common problems when testing algorithms is the lack of a dataset that challenges a specific aspect of the algorithm. Let us take standard k-means which works well on linearly separable clusters, but how will it work on linearly non-separable clusters? To get insight we need a higher-D dataset that has these conditions. Now the challenge is where to get a high-D dataset with multiple clusters that are not linearly separable, even in 4D. We can quickly use our interface to generate such a 4D dataset

We aim to create a dataset with two such clusters spanning the entire 4D space – one of these has an 'L' shaped structure on the *x1-x2* plane while the other has the same structure on the *x2-x3* plane. This guarantees the linear non-separability we strive for. Similar to Figure 10, we decompose the two 4D clusters into four sub-clusters for ease of manipulation. The generation process of the first cluster is exactly the same as in Figure 10. For the second cluster, we initially paint on the *x2-x3* plane, and also go through the sculpting procedure to make sure each sub-cluster only span one dimension. Figure 12a, b show the initially painted distributions, while Figure 12c, d show

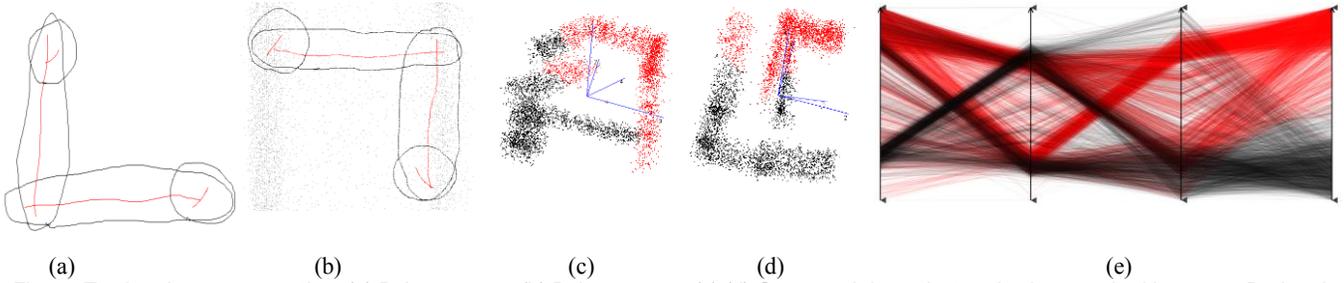

(a) (b) (c) (d) (e)

Fig.12. Testing dataset generation. (a) Paint on *x1-x2* (b) Paint on *x3-x4* (c) (d) Generated data, clustered using standard k-means. Red and black are two clusters. (e) Generated data on parallel coordinates,

the two clusters as scatterplots. Figure 12b shows those points generated from the first distribution painting as background gray points.

We then performed standard k-means clustering on the generated dataset, and we obtained two clusters colored red and black (Figure 12c, d). From the parallel coordinate's view of this dataset (Figure 12e) we observe that the standard k-means clustering mainly operated on the 4[th] dimension and that the resulting two clusters are not accurate. We now know that a more sophisticated clustering algorithm, such as a kernel method, must be used.

### 4.2.6 Existing Data Editing

Our proposed system also allows users to modify an existing data set for better algorithm testing and data analysis purposes.

One of the most encountered problems in data analysis is the existence of outliers. Outliers may impair the ability to discover underling patterns or finding meaningful trends. One may use outlier detection algorithms to remove them, but different algorithms are based on different definitions (density based or distance based) of outliers and may not fit into a particular need. Hence, to be able to visually detect outliers and subsequently remove these outliers would be more suitable in the general case.

Let us take the housing data set [24] as an instance. A user might want to determine the relationship between the 'average number of rooms per dwelling' and the '% lower status of the population'. After plotting the data against these two attributes (Figure 13a), we see that most of the points are concentrated on the main cluster but outliers (Figure 13a, red points) exist. This makes further analysis hard, and simply including all the points in the analysis would introduce errors. Some methods might be able to adjust to the shape of the data, but they often bring the risk of over fitting. Pre-processing the data to remove unwanted points would be very useful. The user could employ general outlier detection algorithms but since this is a high-D data set, the importance of these two attributes might be reduced. In addition, these algorithms often require a pre-setting of some parameters or thresholds which can lead to erroneous results. Our system would come in handy here. The user could navigate to the desired view and visually detect outliers according to their own understanding and need, brushing away unwanted points (Figure 13b). Using the remaining points for further analysis would then yield more robust results.

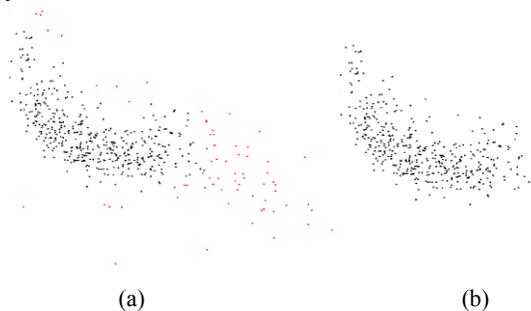

(a) (b)

Fig. 13. Editing an existing data set. (a) Red points are outliers (b) Outliers removed

## 5 IMPLEMENTATION AND USER STUDY

We implemented the parallel coordinates interface in Processing [18], an environment and programming language based on Java. Dataset generation with up to 10 clusters and 1000 samples per cluster runs at interactive speed on a 2.4 GHz Intel Core i5 computer with 4 GB of RAM. The user interface for the Touchpad[N-D] has been implemented in C# and runs on a 2.8 GHz Intel Core i7 computer with 12GB of RAM.

We have also conducted an informal user study to test our interfaces. Eight first-time users were given a brief demonstration of all features of the first user interface, a five-minute tutorial under our guidance, and a task to generate a 6D dataset of 1000 samples. None of the users had prior knowledge about parallel coordinates. One user had generated multivariate datasets before and commented that our user interface could be useful. The feedbacks were generally positive. Some users had difficulty understanding the effect of PDF sketches so they used data connection quadrilaterals instead. The results are not exactly the same but visually similar as shown in Figure 14. Since users could not undo the drawn PDF profiles and data connection quadrilaterals, they reset the sketchpad a few times. On average, users reset 1.5 times and finished the task in 2 minute and 32 seconds of the last attempt.

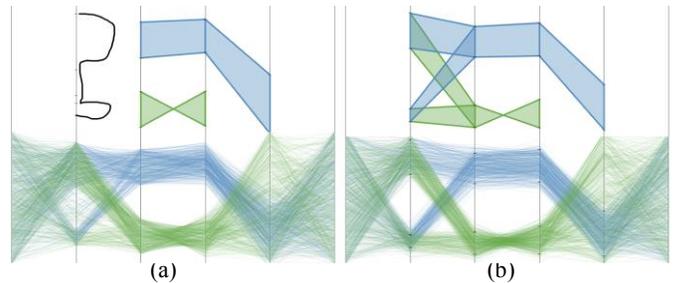

(a) (b)

Fig. 14. Two results by two different users from the user study

## 6 CONCLUSIONS AND FUTURE WORK

We believe that our SketchPad[N-D] can be applied to other visualizations in general. The two designs we proposed both have their strengths and weaknesses. Parallel coordinates show a better overview of the data and the sketching tool is quite easy to interact with. The scatterplot interface, on the other hand, provides a better sense for a structure's distribution but the navigation with the polygon is a bit more abstract. Since our input devices and visualizations all use 2D paradigms, the sketch-based user interfaces are easy to use even for novice users who can casually draw a napkin sketch to specify visual properties of a dataset. Still, we would like to add curve-drawing or line-drawing widgets to facilitate the sketching, as well as better support for categorical data.

### ACKNOWLEDGEMENTS

This work was partly supported by NSF grant IIS-1117132, by the Ministry of Korea Knowledge Economy, Chulalongkorn University, and Fulbright International Science & Technology Awards.

## REFERENCES

[1] G. Albuquerque, T. Löwe, and M. Magnor, "Synthetic generation of high-dimensional datasets," *IEEE Trans on Visualization and Computer Graphics,* 17(12):2317-24, 2011.

[2] E. Amorim, E. Brazil, J. Daniels, P. Joia, L. Nonato, M. Sousa, "iLAMP: Exploring high-dimensional spacing through backward multidimensional projection," *IEEE VAST*, pp. 53-62, 2012.

[3] S. Bremm, T. von Landesberger, M. Heß, D. Fellner, "PCDC - On the Highway to Data - A Tool for the Fast Generation of Large Synthetic Data Sets," *EuroVis Workshop on Visual Analytics*, pp. 7-11, 2012.

[4] J. Browne, B. Lee, S. Carpendale, and N. Riche, "Data Analysis on Interactive Whiteboards through Sketch-based Interaction," *ACM Proc on Interactive Tabletops and Surfaces*, pp. 13-16, 2011.

[5] W. Chao, "NapkinVis: Rapid Pen-Centric Authoring of Improvisational Visualizations," *IEEE Infovis,* poster session, 2010.

[6] W. Chao, "Poster : Rapid Pen-Centric Authoring of Improvisational Visualizations with NapkinVis," *Computing Systems*, 16(6), 2010.

[7] R. DeMillo and A. Offutt, "Constraint-Based Automatic Test Data Generation," *IEEE Trans on Software Engineering*, 17(9):900-910, 1997.

[8] A. Forsberg, M. Dieterich, and R. Zeleznik, "The Music Notepad," *Proc of UIST, ACM SIGGRAPH*, pp.203-210, 1998.

[9] J. Hartigan, "Printer graphics for clustering," *Journal of Statistical Computation and Simulation*, 4(3):187-213, 1975.

[10] T. Igarashi, S. Matsuoka, and H. Tanaka, "Teddy: A Sketching Interface for 3D Freeform Design," *ACM SIGGRAPH*, pp. 409-416, 1999.

[11] A. Inselberg, "The plane with parallel coordinates," *The Visual Computer*, 1(4): 69-91, 1985.

[12] K. Jennings, "Hyperscore : A Graphical Sketchpad for Novice Composers," IEEE *Computer Graphics and Applications,* 24(1):50-54, 2004.

[13] J. LaViola and R. Zeleznik, "MathPad2: A system for the creation and exploration of mathematical sketches," *ACM Trans on Graphics*, 23(3):432-440, 2004.

[14] M. Meyer, H. Lee, A. Barr, M. Desbrun, "Generalized barycentric coordinateson irregular polygons", *Graphics Tools*, 7(1):13-22, 2002.

[15] C. Michael, G. McGraw, M. Schatz, R. Walton, R. Res, and V. Sterling, "Genetic Algorithms for Dynamic Test Data Generation," *Automated Software Engineering*, pp. 307-308, 1997.

[16] J. Nam, K. Mueller, "TripAdvisorN-D: A Tourism-Inspired High-Dimensional Space Exploration Framework with Overview and Detail," *IEEE Trans on Visualization and Computer Graphics,* 19(2):291-305, 2013.

[17] R. Pargas, M. Harrold, and R. Peck, "Test-Data Generation Using Genetic Algorithms," *Software Testing, Verification and Reliability*, pp. 41-48, 1999.

[18] C. Reas and B. Fry, "Processing: a learning environment for creating interactive Web graphics," *Proc of the SIGGRAPH conference on Web graphics*, pp. 1, 2003.

[19] P. Ruchikachorn, B. Wang, K. Mueller, "SketchPad N-D: An Interface for High-Dimensional Dataset Generation and Editing," *IEEE Visualization,* poster session, 2012.

[20] D. Tenneson and S. Becker, "ChemPad: Generating 3D Molecules From 2D Sketches," *ACM SIGGRAPH*, poster, pp. 11, 2005.

[21] R. Zeleznik, K. Herndon, and J. F. Hughes, "SKETCH: An Interface for Sketching 3D Scenes," *Proc of SIGGRAPH*, pp. 163-170, 1996.

[22] R. Zeleznik, T. Miller, and C. Li, "MathPaper: Mathematical sketching with fluid support for interactive computation," *Smart Graphics*, pp. 1-12, 2008.

[23] T. Baudel, "From information visualization to direct manipulation: extending a generic visualization framework for the interactive editing of large datasets," *Proc of the ACM symposium on User interface software and technology*, pp 67-76, 2006.

[24] http://archive.ics.uci.edu/ml/datasets/Housing

[25] https://code.google.com/p/google-refine/